\documentclass[aps,prc,preprint,superscriptaddress,floatfix,showpacs]{revtex4-1}
\usepackage{epsfig}
\usepackage{dcolumn}
\usepackage{bm}
\usepackage{amssymb}
\usepackage{amsmath}

\begin{document}
\title{Signatures of  thermalized charm quarks in all charged flow observables}

\author{J. Noronha-Hostler}
\affiliation{\small{ Department of Physics and Astronomy, Rutgers University, Piscataway, NJ USA 08854}}
\author{C. Ratti}
\affiliation{\small{ Department of Physics, University of Houston, Houston, TX, USA  77204}}

\date{\today}
\begin{abstract}
A long standing question in the field of heavy-ion collisions is whether charm quarks are thermalized within the Quark Gluon Plasma.  In recent years, progress in lattice QCD simulations has led to reliable results for the equation of state of a system of 2+1 flavors (up, down, and strange) and 2+1+1 flavors (up, down, strange, and charm).  We find that the equation of state  strongly affects differential flow harmonics and a preference is seen for thermalized charm quarks at the LHC. Predictions are also made for the event-plane correlations at RHIC AuAu $\sqrt{s_{NN}}=200$ GeV collisions, and the scaling of differential flow observables and factorization breaking for all charged particles at LHC PbPb $\sqrt{s_{NN}}=5.02$ TeV collisions compared to LHC XeXe $\sqrt{s_{NN}}=5.44$ TeV collisions, which could be useful in answering the question: are charm quarks thermalized? 
\end{abstract}

\maketitle

Shortly after the Big Bang, a state of matter known as the Quark Gluon Plasma (QGP) filled the entire universe \cite{Collins:1974ky}.  Since the early 2000's, the QGP has been recreated in the laboratory using high-energy heavy-ion collisions at RHIC and the LHC.  Lattice Quantum Chromodynamics (QCD) calculations have been driving the theoretical understanding of the QGP \cite{Philipsen:2012nu,Ratti:2018ksb}. They provide the equation of state (EoS) from first principles at zero and small baryon densities.  Lattice QCD demonstrated that the early universe and high-energy heavy-ion collisions undergo a cross-over phase transition between the QGP phase and a hadron resonance gas \cite{Aoki:2006we}. This could only be determined with fine lattice grids that can be extrapolated to the continuum \cite{Borsanyi:2010bp}.  These first-principle calculations have now progressed to the point that it is possible to directly compare the EoS of 2+1 flavors (up, down, and strange) \cite{Borsanyi:2013bia,Bazavov:2014pvz} vs. 2+1+1 flavors (up, down, strange, and charm) \cite{Borsanyi:2016ksw}.  A system with 2+1+1 dynamical flavors implies that charm quarks are thermalized whereas a system with 2+1 dynamical flavor assumes thermalization only for the light and strange quarks. 

While it is believed that charm quarks should be thermalized in the Early Universe in the quark epoch \cite{Borsanyi:2016ksw} due to its comparatively larger size and slower expansion, heavy-ion collisions produce only a handful of charm quarks per collision so it is not clear whether they are thermalized in this case \cite{Moore:2004tg,Mannarelli:2005pz,Rapp:2009my,Cao:2011et,Adhya:2014nza,Nahrgang:2016lst,Mukherjee:2015mxc,Scardina:2017ipo,Prado:2017fjn,Xu:2017obm}. Many of the approaches attempting to answer this question have looked at measurements of the nuclear modification of heavy quarks and their flow harmonics, which are plagued by low statistics. Previous approaches have pointed to the large elliptical flow of charm \cite{Pal:2005xy,Adamczyk:2017xur,TheATLAScollaboration:2015bqp,Adam:2016ssk,Sirunyan:2017plt,Sirunyan:2018toe}, which may indicate that charm quarks are thermalized \cite{Moore:2004tg}.  However, other works have discussed the very close proximity of charm quarks in phase space, which would make it difficult for them to fully thermalize with the medium \cite{Martinez:2018ygo}.  In this letter we propose a radically new method for determining if charm quarks are thermalized, that can be easily tested using all charged particle flow observables.

Until now, the most successful tool to study the dynamical properties of the QGP has been event-by-event relativistic viscous hydrodynamics \cite{Gale:2012rq,Niemi:2012aj,Niemi:2015voa,Noronha-Hostler:2015uye,Niemi:2015qia,Giacalone:2016eyu,Giacalone:2016afq,Noronha-Hostler:2016eow,Betz:2016ayq,Bernhard:2016tnd,McDonald:2016vlt,Ke:2016jrd,Zhao:2017yhj,Alba:2017hhe,Giacalone:2017uqx}. The exception has been that, at LHC run 2, hydrodynamic predictions yield a poor description of $v_n(p_T)$ in the soft sector ($p_T<3$ GeV) \cite{Acharya:2018lmh}. 

 The EoS from Lattice QCD can be fed directly into the hydrodynamic models in order to make direct comparisons to experimental data. A recent Bayesian analysis using standard hydrodynamical observables (that require low statistics) has confirmed that the Lattice QCD EoS gives approximately the right form in order to reproduce the experimental data \cite{Pratt:2015zsa}.  However, most of the common observables are not particularly sensitive to the EoS \cite{Huovinen:2009yb,Moreland:2015dvc,Bernhard:2016tnd,Alba:2017hhe} and while $v_1$ is known to be sensitive to the EoS \cite{Stoecker:2004qu,PhysRevC.90.014903,Xing:2010zzb,PhysRevC.91.024915,Bravina:2016dag,Nara:2016hbg}, it is influenced by a complicated mix of eccentricities \cite{Gardim:2014tya}. Besides, momentum conservation must also be considered \cite{Gardim:2011qn}, making it a poor choice for constraining the EoS. In this letter, we point out the high sensitivity of event plane correlations, differential flow, and factorization breaking to the assumptions regarding the flavor number of thermalized quarks in the EoS. We find that the EoS with thermalized charm quarks significantly improves the fit to $v_n(p_T)$ at LHC run 2. The scaling behavior from PbPb to XeXe is also significantly affected by the EoS so we make predictions that, if confirmed, could be a further proof that charm quarks are thermalized at the LHC.

\begin{figure}[ht]
\centering
\includegraphics[width=0.5\textwidth]{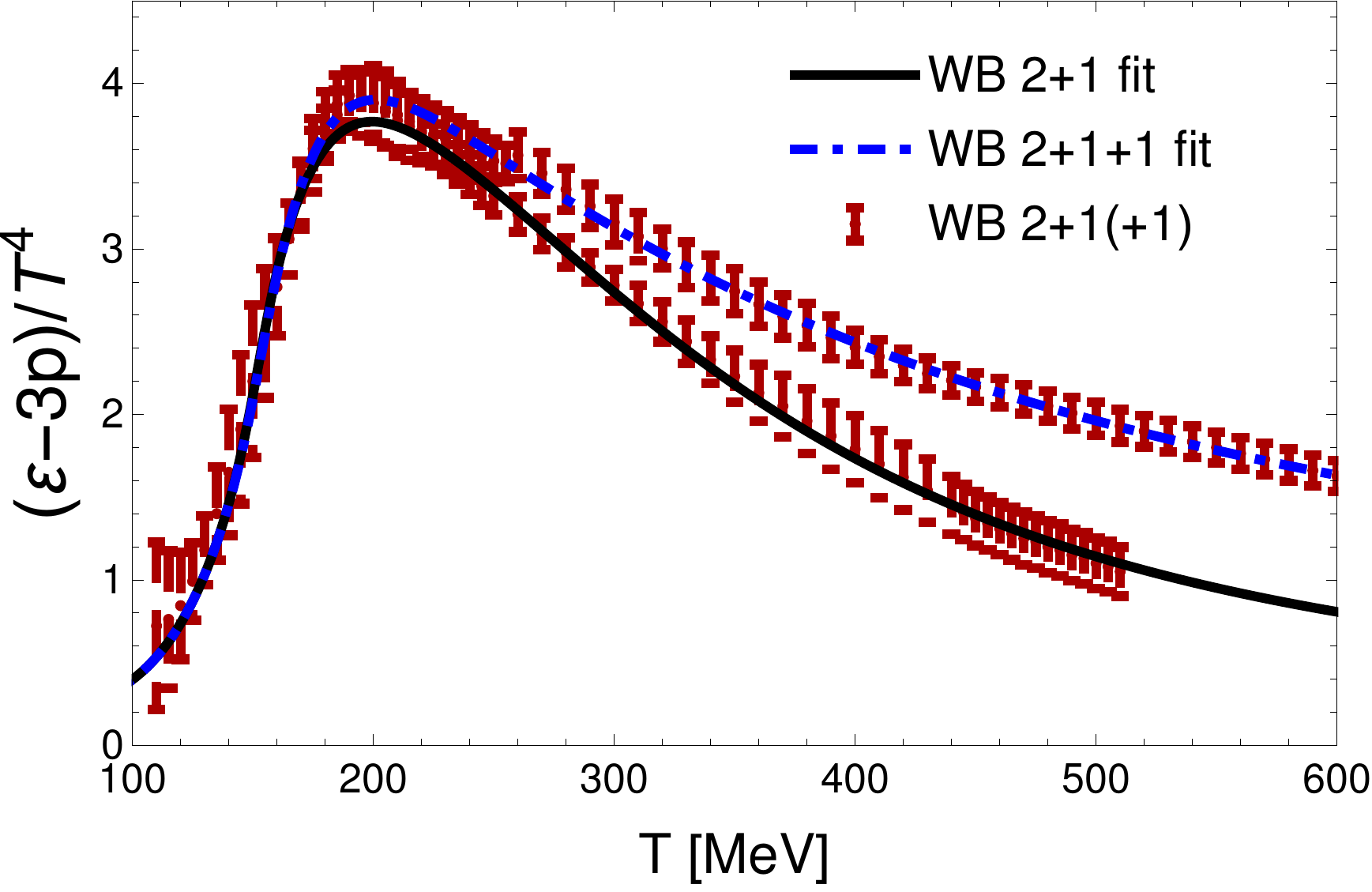} 
\caption{(Color online) Trace anomaly of the 2+1-  \cite{Borsanyi:2013bia} vs. the  2+1+1-flavor EoS from the quark epoch in the Early universe from Ref. \cite{Borsanyi:2016ksw} }
\label{fig:trace}
\end{figure}

Because of the large mass of the charm quark ($m_c\sim 1.3$ GeV), it is commonly assumed that charm quarks would only be influential at temperatures close to $T\sim 1.3$ GeV.  Contrary to that assumption, in \cite{Borsanyi:2016ksw} it was shown that thermalized charm quarks could influence the thermodynamic properties at temperature even below $T\lesssim 300$ MeV, which are reached by both RHIC and LHC energies, as shown in Fig.\ \ref{fig:trace}.  
In Fig.\ \ref{fig:trace} we show the parameterizations of the trace anomaly for the two different equations of state used in this letter, compared to the latest Wuppertal-Budapest collaboration data for 2+1 flavors \cite{Borsanyi:2013bia,Bluhm:2013yga} and 2+1+1 flavors \cite{Borsanyi:2016ksw}. We complement the lattice QCD results at low temperatures through a smooth merging to the Hadron Resonance Gas model, for which only the most up-to-date particle resonance lists are used throughout the paper, the so called PDG2016+ from \cite{Alba:2017mqu,Alba:2017hhe}. Based on our two parameterized EoS we see a $14\%$, $40\%$, $70\%$, and $102\%$ increase in the trace anomaly of the 2+1+1-flavor EoS over the 2+1-flavor EoS at temperatures $T=300$, 400, 500, and 600 MeV, respectively. This demonstrates that at LHC, where a maximum temperature of $T_{max}\sim 600$ MeV is reached, it is likely that the influence of thermalized charm quarks can be measured.

We use event-by-event fluctuating initial conditions generated from the TRENTO model \cite{Moreland:2014oya}  tuned to IP-Glasma \cite{Gale:2012rq} i.e. $p=1$, $k=1.6$, and $\sigma=0.51$, which are known to successfully reproduce the experimental data \cite{Giacalone:2017uqx,Alba:2017hhe}. A very fine initial grid size of the initial conditions is set to $dx=dy=0.06$ fm at AuAu $200$ GeV and $dx=dy=0.05$ fm at PbPb 5.02 TeV and XeXe 5.44 TeV and 35,000-50,000 initial conditions are generated at each energy. Here only deformed XeXe events are used \cite{Giacalone:2017dud}.   We use the relativistic viscous hydrodynamics code, v-USPhydro, to evolve the initial conditions on an event-by-event basis.  Hydrodynamics is switched on at $\tau_0=0.6$ fm for both RHIC and LHC, where it is evolved with the smoothing parameter $h=0.3$ fm (see \cite{Noronha-Hostler:2013gga,Noronha-Hostler:2014dqa,Noronha-Hostler:2015coa} for more details). At the switching temperature $T_{SW}=150$ MeV the fluid is hadronized using Cooper-Frye \cite{Cooper:1974mv,Anderlik:1998cb}.  The resonance decays are described using an adapted version of AZHYDRO \cite{Kolb:2000sd,Kolb:2002ve,Kolb:2003dz} with the full PDG2016+ particle list.  Further details on the inclusion of these new resonances and their decay channels can be found in \cite{Alba:2017hhe}.  In the following, if high statistics is needed for a specific observable, we always show the effect of our sample size via jackknife resampling. 

For the time being, no bulk viscosity is considered, which we would expect to alter our $\langle p_T\rangle$ results and possibly  to the  slope of $v_n(p_T)$ \cite{Noronha-Hostler:2013gga,Noronha-Hostler:2014dqa,Ryu:2015vwa}. However, we would not expect a need for a large $\zeta/s$ in our set-up because we have a reasonable fit to $\langle p_T\rangle$ already \cite{Alba:2017hhe}. Additionally, hadronic transport (such as UrQMD) is not considered because this would require adaptation both of our particle resonance list to only include 2 body interactions as well as an adaptation of UrQMD itself to include these new states, which is outside of the scope of this paper. At RHIC $\eta/s=0.05$ for both EoS and at PbPb 5.02 TeV and XeXe 5.44 TeV $\eta/s=0.047$ for EoS 2+1 and $\eta/s=0.04$ for EoS 2+1+1.  The values of $\eta/s$ were chosen by comparison of integrated $v_2\{2\}$ and $v_3\{2\}$ to experimental, as discussed in \cite{Alba:2017hhe}.  All results shown here are predictions  entirely based upon this previous set-up established in \cite{Alba:2017hhe}. Differences between the spectra and integrated flow harmonics are negligible  at both PbPb 5.02 TeV and XeXe 5.44 TeV. The $\langle p_T\rangle$ of protons is $\sim 3\%$ higher for the 2+1-flavor EoS at XeXe 5.44 TeV and $\sim 3-5\%$ higher for the 2+1-flavor EoS at PbPb 5.02 TeV. The difference in $\langle p_T\rangle$ can be explained by the slightly larger $\eta/s$ for 2+1-flavors, as a larger $\eta/s$ has been previously found to increase $\langle p_T\rangle$ \cite{Luzum:2008cw}.

\begin{figure}[ht]
\centering
\includegraphics[width=0.75\textwidth]{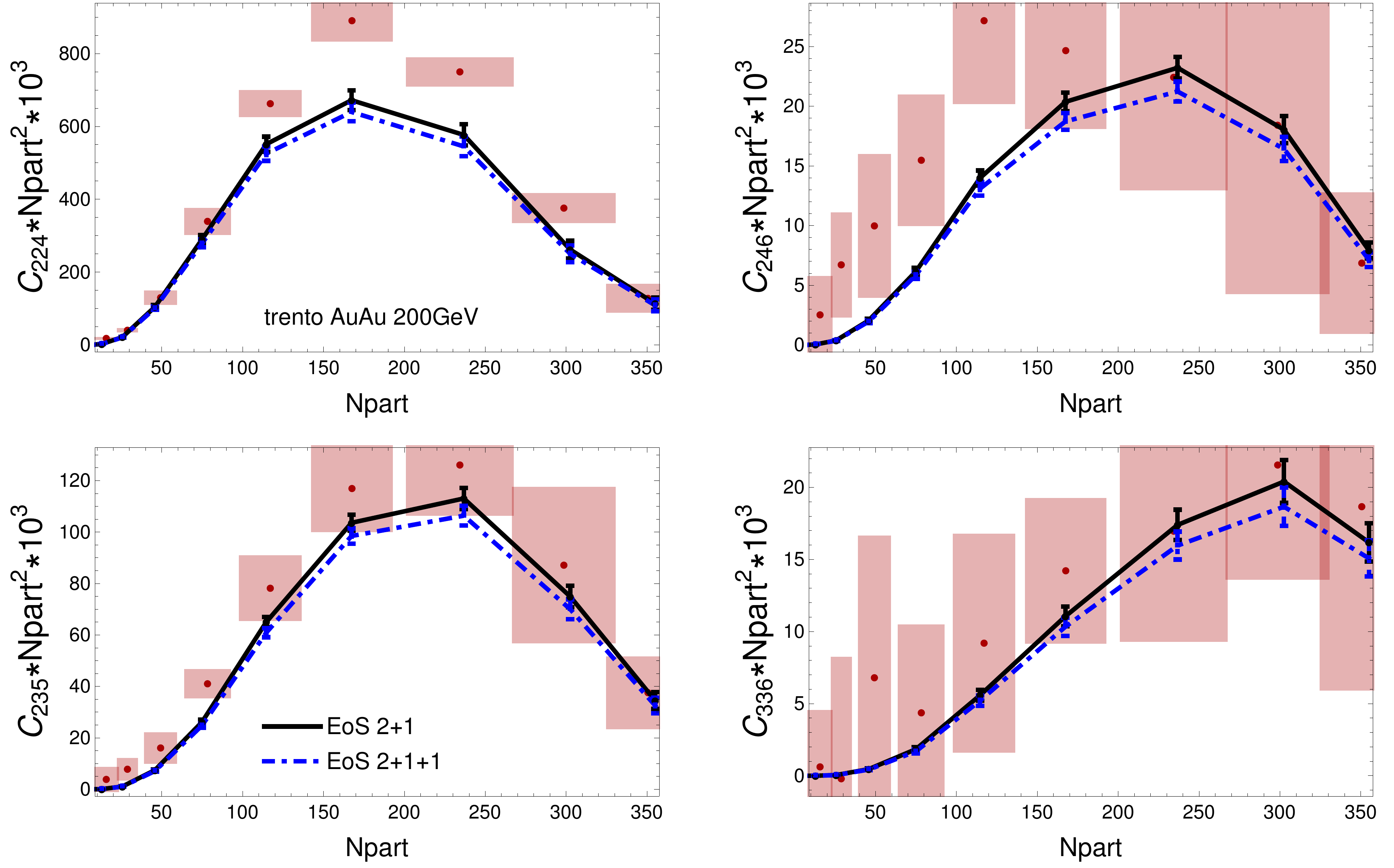}
\caption{(Color online) $Npart^2*C_{n,m,n+m}*10^3$ at AuAu $\sqrt{s_{NN}}=200$ GeV  for charged particles from STAR \cite{Adamczyk:2017byf} compared to the 2+1-  and the 2+1+1-flavor EoS . }
\label{fig:AuAu}
\end{figure}
The primary difference between the 2+1- and 2+1+1-flavor equation of state appears at high temperatures, so we do not expect a significant difference in flow observables at RHIC. Indeed, for integrated flow observables in \cite{Alba:2017hhe} no differences were seen between 2+1- and 2+1+1-flavors, regardless of energy and system size. Additionally, we do not observe clear differences in the differential flow observables nor in the factorization break at RHIC (see Appendix \ref{sec:rhic}). The only observables that do appear to have any sensitivity to the difference between the two EoS's are the event plane correlations measured by STAR \cite{Adamczyk:2017byf}. Event-plane correlations were first measured experimentally by the ATLAS collaboration \cite{Aad:2014fla}, that also suggested the alternative normalization \cite{Acharya:2017zfg}.  In \cite{Adamczyk:2017byf} only the numerator is measured, namely
\begin{equation}
C_{n,m,n+m}=\langle v_m v_n v_{m+n} \cos \left(m\Psi_m+n\Psi_n-(m+n)\Psi_{m+n}\right) \rangle,
\end{equation}
and our prediction from Fig.\ \ref{fig:AuAu} using the 2+1-flavor EoS was shown to be the best fit in \cite{Adamczyk:2017byf}. 
In Fig. \ref{fig:AuAu}, a comparison of $C_{n,m,n+m}$ calculated using the two equations of state is shown: the 2+1-flavor EoS shows a stronger correlation between event plane angles, which yields a slightly better agreement to the experimental data. If infinite statistics were possible both in theory and experiment, this might be a distinguishing observable for the thermalization of charm quarks at RHIC. However, at the moment the experimental error-bars are unfortunately quite large. For this reason, we explore flow observables at the LHC.

\begin{figure}
\centering
\begin{tabular}{c c}
\includegraphics[width=0.5\textwidth]{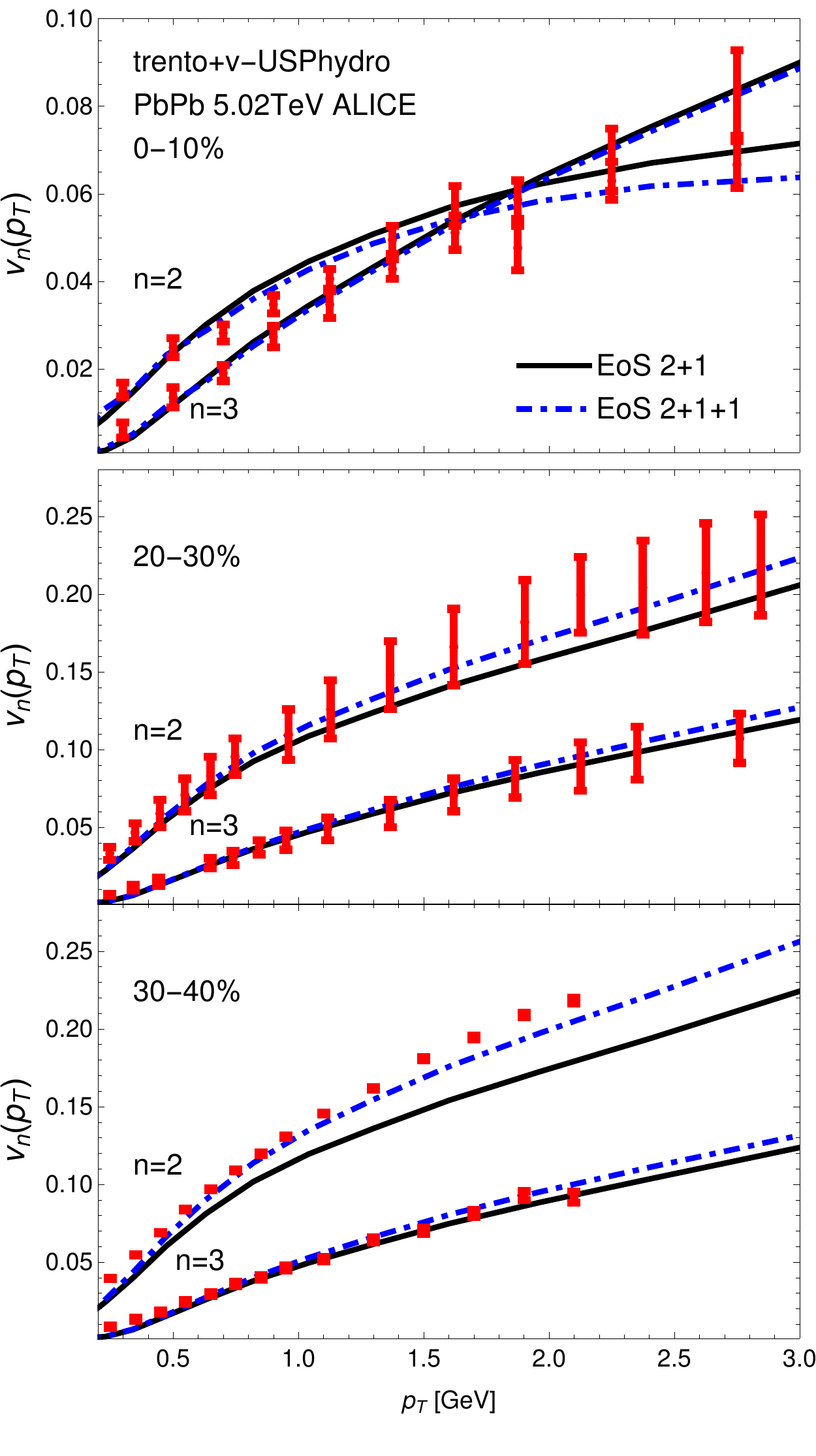} & \includegraphics[width=0.5\textwidth]{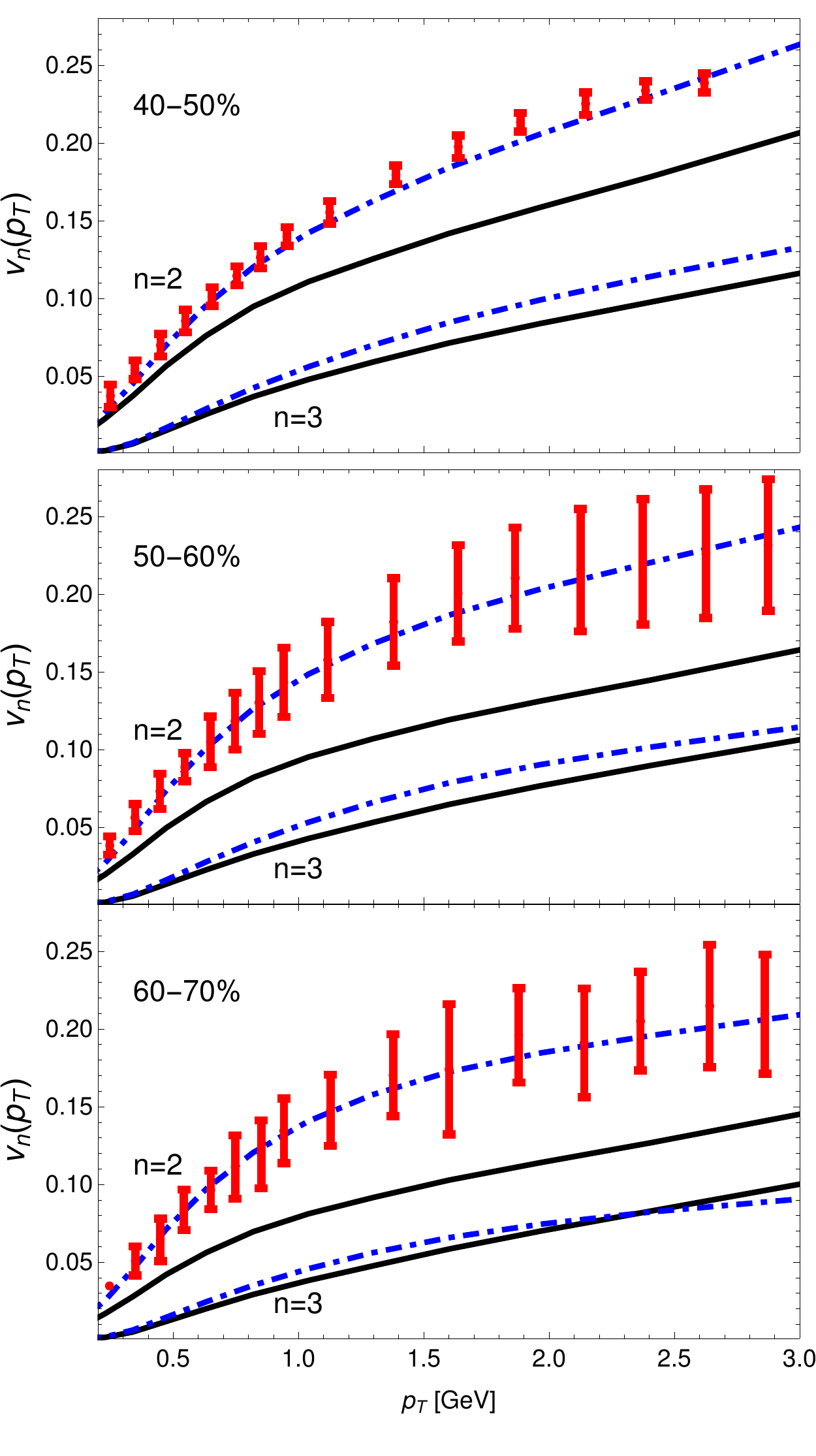}
\end{tabular}
\caption{(Color online) $v_n(p_T)$ calculations at PbPb $5.02$ TeV for charged particles for the 2+1-flavor EoS from Ref. \cite{Borsanyi:2013bia} and the 2+1+1-flavor EoS  \cite{Borsanyi:2016ksw} compared to the experimental data from ALICE \cite{Adam:2016izf,Acharya:2018lmh}. }
\label{fig:vnptlhc}
\end{figure}

In Fig.\ \ref{fig:vnptlhc} the differential flow of $v_n(p_T)$ for n=2,3 is shown.  Notice that all theory calculations use the scalar product \cite{Bilandzic:2010jr,Bilandzic:2013kga}
\begin{equation}
v_n(p_T)=\frac{\langle v_2 v_2(p_T)\cos\left[n\left(\Psi_n -\Psi_n(p_T)\right)\right]\rangle}{\sqrt{\langle v_2^2\rangle}}
\end{equation}
with multiplicity weighing and centrality rebinning \cite{Noronha-Hostler:2016eow,Betz:2016ayq,Gardim:2016nrr}. Comparing to the experimental data from ALICE \cite{Acharya:2018lmh}, we find an excellent agreement for the 2+1+1-flavor EoS, while the 2+1-flavor EoS fails in peripheral collisions. Part of the difference can be attributed to the $15\%$ smaller viscosity needed for the 2+1+1-flavor EoS to fit integrated flow harmonics and because nonlinear effects are largest in peripheral collisions \cite{Noronha-Hostler:2015dbi}.  The 2+1+1-flavor EoS consistently produces a larger $v_2$ that matches the ALICE data across a range of centrality classes in peripheral collisions, which makes a strong case for the thermalization of charm quarks.

\begin{figure}
\centering
\includegraphics[width=0.75\textwidth]{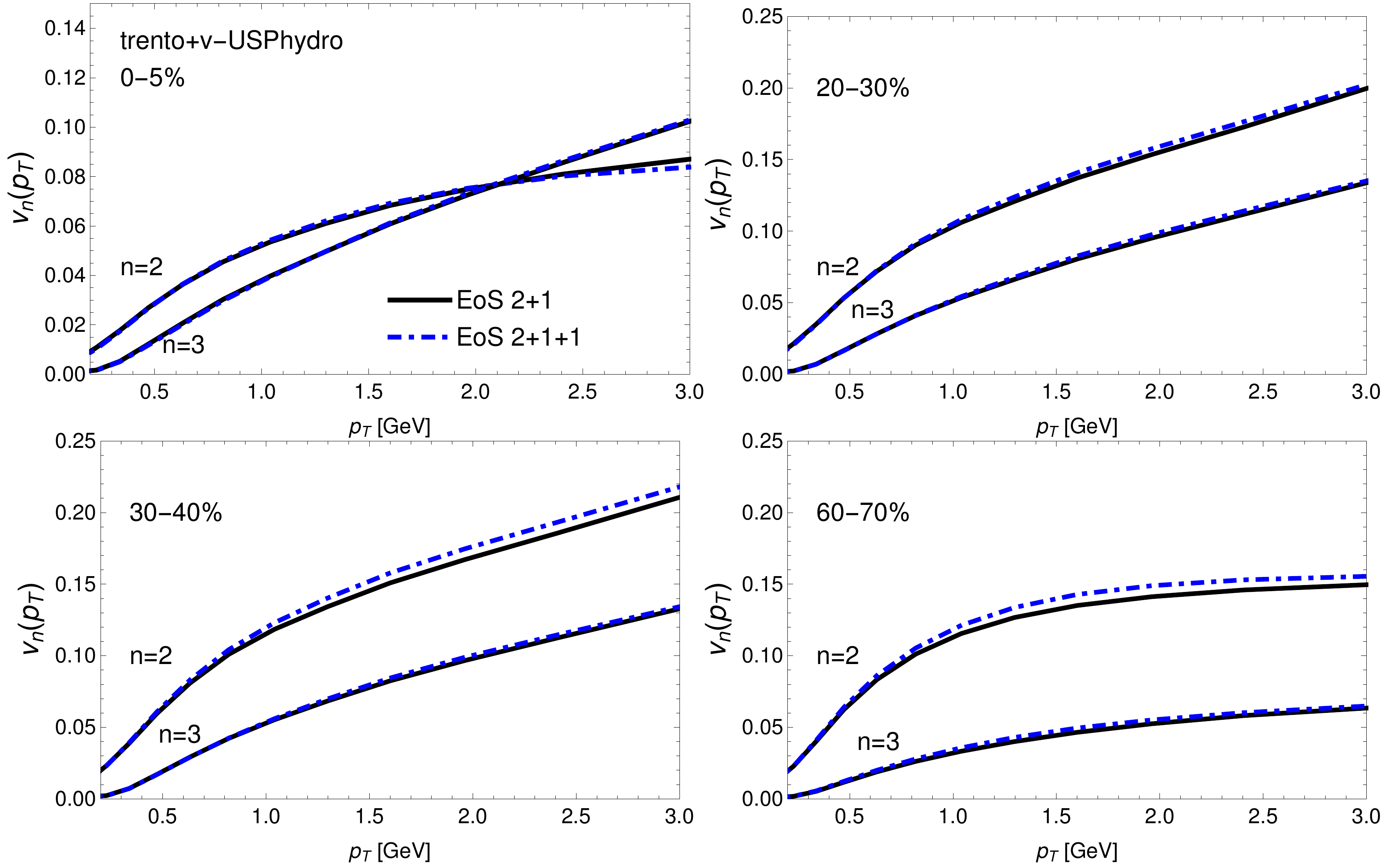}
\caption{(Color online) $v_n(p_T)$ for n=2,3 calculations at XeXe $5.44$ TeV for charged particles for the 2+1- vs 2+1+1-flavor EoSs. }
\label{fig:xe}
\end{figure}

Because $v_n(p_T)$ in peripheral collisions appears to be sensitive to the EoS, we study the effects of system size scaling.  XeXe collisions were ran this year at nearly the same energy as PbPb collisions (5.44 TeV and 5.02TeV, respectively).  In Fig.\ \ref{fig:xe}, $v_n(p_T)$ in XeXe collisions is shown, calculated using the two EoS. Unlike in PbPb collisions, XeXe collisions predict nearly identical results for the differential flow harmonics when varying the EoS. 
Comparing Figs.\ \ref{fig:vnptlhc}-\ref{fig:xe} we expect significantly different results for the 2+1- vs. 2+1+1-flavor equations of state when scaling between PbPb and XeXe collisions.  A simple test of the thermalization of charm quarks is to compare $v_2(p_T)$ in PbPb to XeXe. If a suppression of $v_2(p_T)$ is seen in XeXe, it appears that charm quarks are thermalized.

\begin{figure}
\centering
\includegraphics[width=0.5\textwidth]{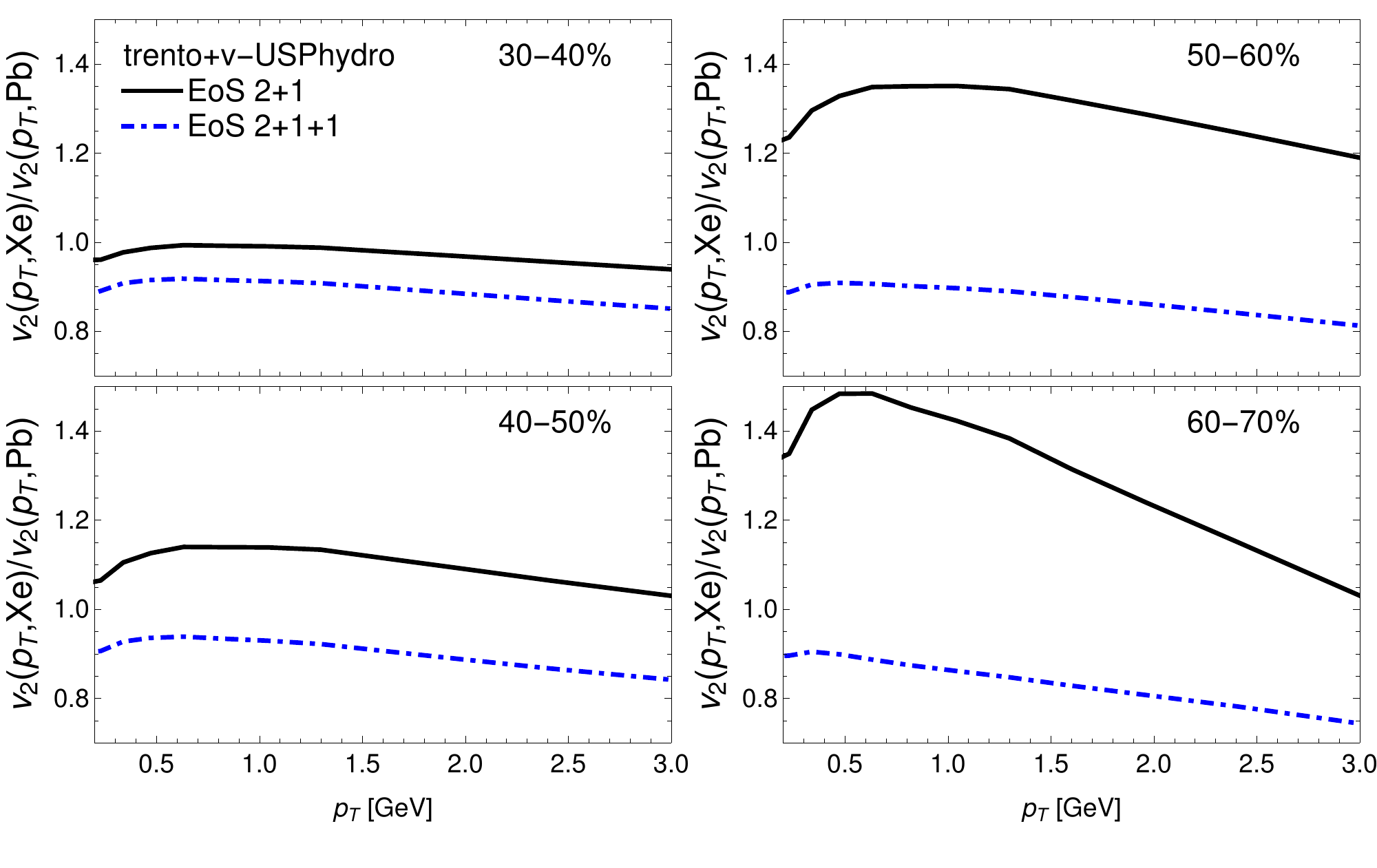}
\caption{(Color online) $v_2(p_T)$ at PbPb $5.02$ TeV divided by $v_2(p_T)$ at XeXe $5.44$ TeV for charged particles for the 2+1- vs.  2+1+1-flavor Equation of State. }
\label{fig:rat}
\end{figure}

To analyze the differences between the larger system size in PbPb collisions vs. XeXe collisions, we also plot the ratio of the differential elliptical flow at XeXe divided by PbPb collisions, $v_2(p_T,Xe)/v_2(p_T,Pb)$, in Fig.\ \ref{fig:rat}. The ratio has the advantage that some modelling effects my cancel out.   From Figs. \ref{fig:vnptlhc}-\ref{fig:xe}, one can clearly see that in central to mid-central collisions the difference between the two EoS is negligible, so we focus only on the 30-70$\%$ centralities in Fig.\ \ref{fig:rat}.  The splitting is enhanced at more peripheral collisions, wherein the 2+1-flavor EoS predicts an increase in $v_2(p_T,Xe)/v_2(p_T,Pb)$ while the 2+1+1-flavor EoS predicts a decrease in $v_2(p_T,Xe)/v_2(p_T,Pb)$. Because only the 2+1+1-flavor EoS matches experimental data in PbPb collisions, we predict a suppression in $v_2(p_T,Xe)/v_2(p_T,Pb)$, which -if confirmed- points to a scenario where charm quarks are thermalized.  

\begin{figure}[ht]
\centering
\includegraphics[width=0.75\textwidth]{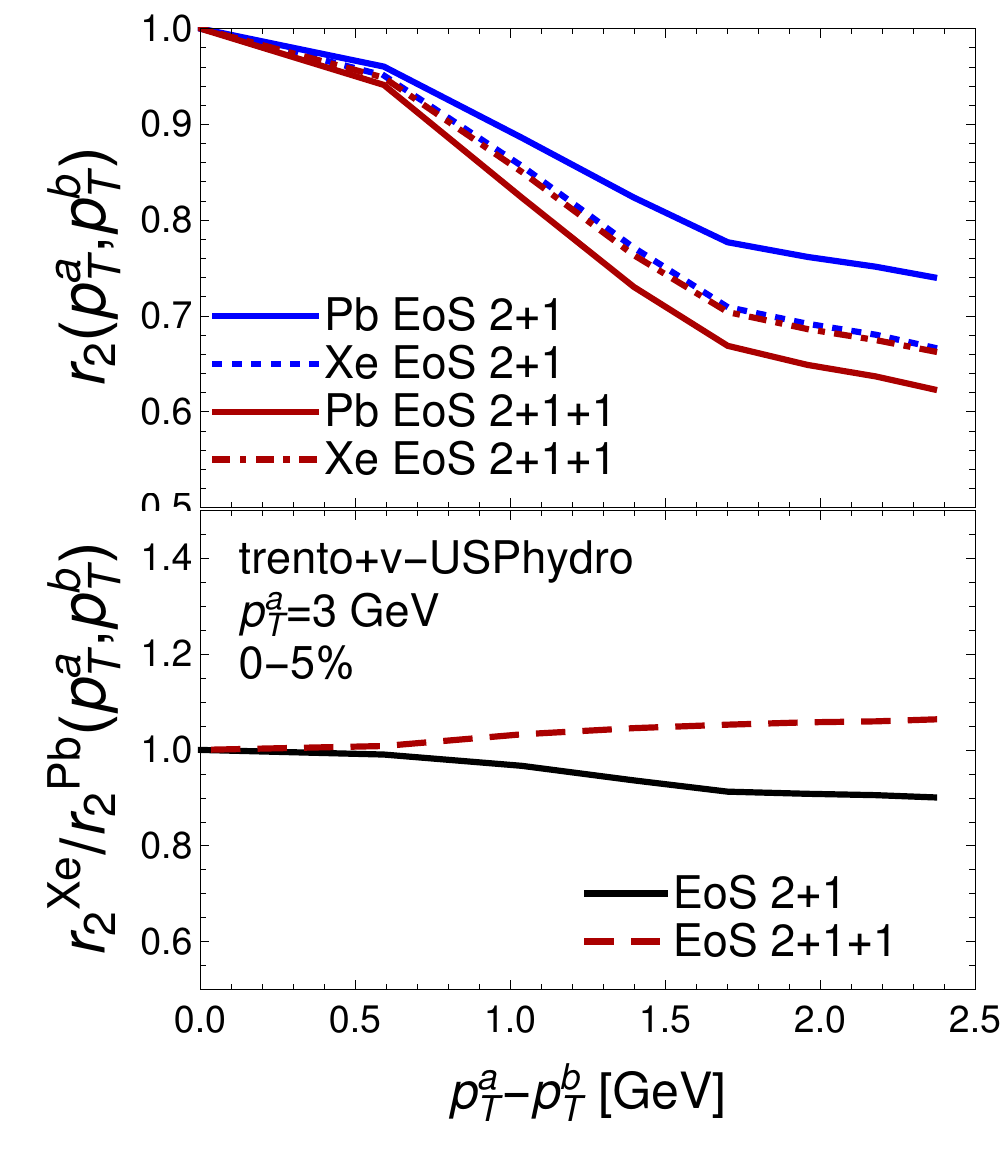}
\caption{(Color online) Upper panel: Factorization breaking  $r_2(p_T^a,p_T^b)$ at PbPb $5.02$ TeV and XeXe $5.44$ TeV for charged particles at $0-5\%$ for the 2+1-flavor EoS from \cite{Borsanyi:2013bia} and the 2+1+1-flavor EoS \cite{Borsanyi:2016ksw}. The lower panel shows our predictions for the ratios of $r_2$ of XeXe divided by PbPb. }
\label{fig:rn05}
\end{figure}

Finally, we study the dependence of the factorization breaking \cite{Gardim:2012im} for $p_T^a=3$ GeV of the elliptical flow  
\begin{equation}
r_2(p_T^a,p_T^b)=\frac{\langle v_2^2(p_T^a) v_2^2(p_T^b)\cos\left[2\left(\Psi_2(p_T^a) -\Psi_2(p_T^b)\right)\right]\rangle}{\sqrt{\langle v_2^2(p_T^a)\rangle\langle v_2^2(p_T^b)\rangle}}
\end{equation}
on the EoS and find that central collisions appear to be strongly sensitive to the EoS in both PbPb and XeXe: we show this result in Fig.\ref{fig:rn05}. Taking the ratio of XeXe $5.44$ TeV over PbPb $5.02$ TeV we find that the 2+1-flavor EoS predicts a decrease in the factorization breaking as the system size decreases, whereas the 2+1+1-flavor EoS predicts a significant increase in $r_2(p_T^a,p_T^b)$ as the system size decreases.  While we show only the results for $p_T^a=3$ GeV, we see this even at smaller $p_T^a$, although for $p_T^a<2$ GeV the effect is small.  Additionally, this splitting of the behavior between XeXe and PbPb collisions for the two different EoS is observed up to $10\%$ centrality, although the difference is the most dramatic in $0-5\%$.

\begin{figure}[ht]
\centering
\includegraphics[width=0.75\textwidth]{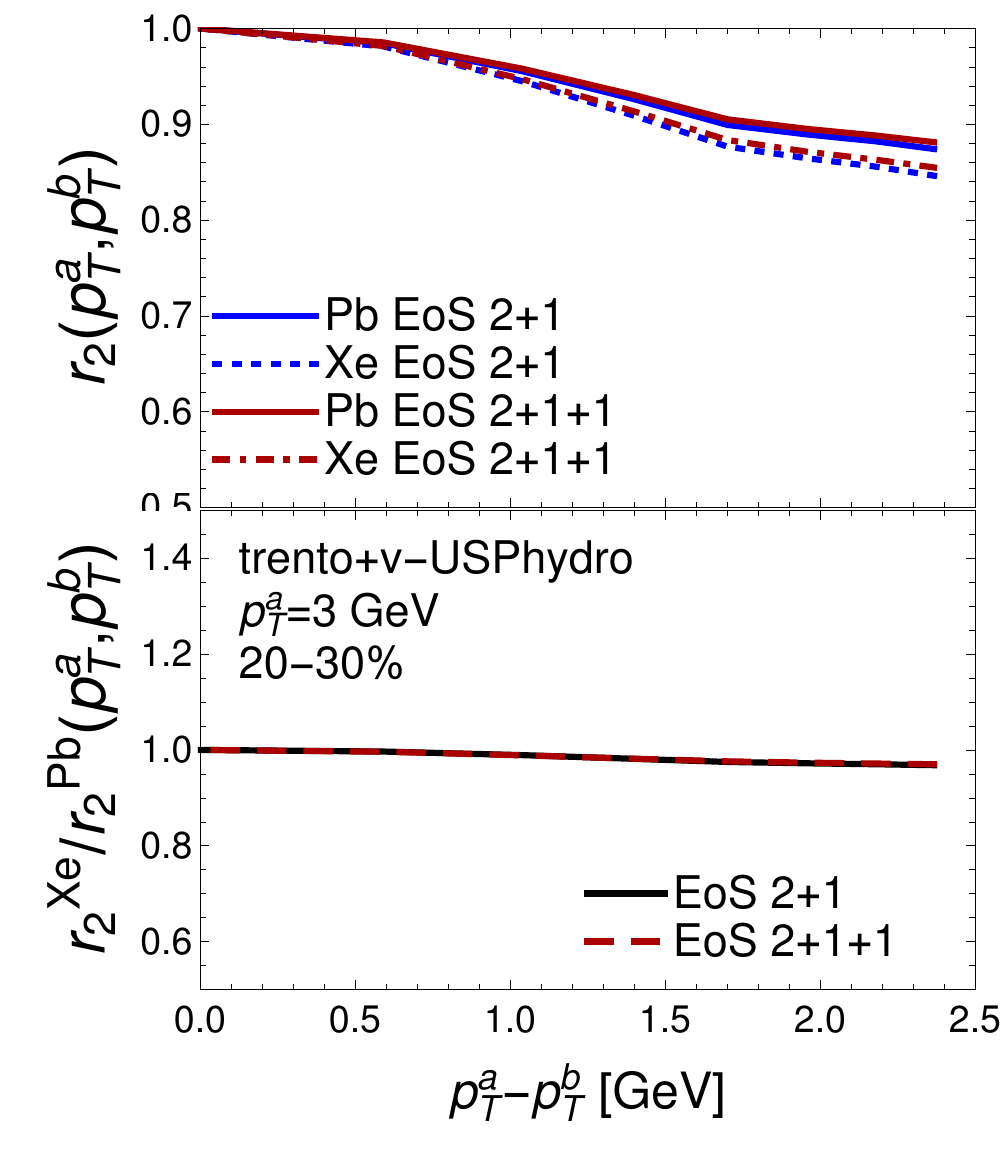}
\caption{(Color online)  Factorization breaking  $r_2(p_T^a,p_T^b)$ at PbPb $5.02$ TeV and XeXe $5.44$ TeV for charged particles at $20-30\%$ for the 2+1-flavor EoS from \cite{Borsanyi:2013bia} and the 2+1+1-flavor EoS \cite{Borsanyi:2016ksw}.  Ratios of $r_2$ of XeXe divided by PbPb predictions are shown on bottom.}
\label{fig:rn25}
\end{figure}

The factorization breaking of mid-central and peripheral collisions have almost no dependence on the EoS, as can be seen for $20-30\%$ centrality in Fig. \ref{fig:rn25}. While the effect on the factorization breaking is interesting and could possibly be used as a secondary constraint, the factorization breaking is also strongly dependent on the initial conditions \cite{Shen:2015qta} while also displays some dependence on the hadronic rescattering \cite{McDonald:2016vlt} and smoothing scale \cite{Gardim:2017ruc} and it is not clear yet how these effects scale with the system size.  However, the sensitivity of a variety of flow observables to the inclusion of thermalized charm quarks in the EoS can only be quantitatively addressed via a global analysis across multiple energies, centrality classes, and observables.

In conclusion, the results of this letter demonstrate an entirely new approach for answering the question of the thermalization of charm quarks in heavy-ion collisions. Unlike previous approaches to study the thermalization of charm quarks that focus on the flow of D mesons, our results are calculated entirely in the soft sector using all charged hadrons.   Using the QCD EoS that includes the thermalization of charm quarks compared to that with only light and strange quarks, we showed previously \cite{Alba:2017hhe} that standard observables such as spectra and integrated flow harmonics see negligible effects from the thermalization of charm quarks. In this work, we motivate our hydrodynamic description using the most state-of-the-art Lattice QCD results for the number of particle resonances, freeze-out temperature, and EoS. Then, we make predictions for observables that do show a sensitivity to the thermalization of charm quarks within the EoS. At RHIC energies, the difference between the two EoS is minimal and the only observables in the soft sector that are affected by charm quarks are the event-plane correlations, although a significant improvement in the experimental error-bars would be needed to answer the question of the thermalization of charm quarks at RHIC.  

At the LHC the highest temperature region of the EoS, that is the most sensitive to the thermalization of charm quarks in the trace anomaly, is probed and the 2+1+1-flavor EoS produces a significantly larger $v_2(p_T)$ in peripheral collisions, which is preferred by the experimental data. The inclusion of thermalized charm quarks predicts a suppression in the ratio $v_2(p_T,Xe)/v_2(p_T,Pb)$ whereas the opposite effect is seen from the 2+1-flavor EoS. If a suppression is experimentally observed in the ratio of $v_2(p_T)$ at XeXe collisions vs. PbPb in peripheral collisions, then this would further build the case for thermalized charm quarks at  the LHC.  Furthermore, because the factorization breaking in central collisions also has clear quantitative differences when one scales from PbPb collisions to XeXe, we are able to make two testable predictions that should further confirm whether charm quarks are thermalized.  We note that, if it appears that the 2+1-flavor EoS is favored, a general statement could be made that within heavy-ion collisions charm quarks are not thermalized, as it is highly unlikely that charm quarks would be thermalized at RHIC. If the 2+1+1-flavor EoS is more favorable at the LHC compared to experimental data, other methods must be used to answer the question whether charm quarks are thermalized also at RHIC. From the currently existing experimental data at the LHC, we see indications that charm quarks are thermalized because of the favored 2+1+1-flavor EoS; to confirm this statement our predictions for the system scale of $v_n(p_T)$ and $r_2$ must be verified.  If the 2+1+1-flavor EoS is confirmed, we will have solid proof that the EoS probed at the highest LHC energies is the same as that in the quark epoch of the Early Universe.

This work raises a number of questions for the future such as what is the effect of the thermalized charm on the differential flow in the hard sector?  Additionally, would a Bayesian analysis also find other new flow observables that could be sensitive to the difference between the 2+1-flavor and 2+1+1-flavor EoS? If charm quarks are thermalized at the LHC, how can that be understood in terms of the charm quark phase-space separation? Finally, are there other flow observables at RHIC that could provide a more conclusive answer at lower energies?

\section{Acknowledgements}

The authors would like to thank Anthony Timmins for his comments and suggestions. 
J.N.H. acknowledges the Office of Advanced Research Computing (OARC) at Rutgers, The State University of New Jersey for providing access to the Amarel cluster and associated research computing resources that have contributed to the results reported here. 
This material is based upon work supported by the National Science Foundation under grants no. PHY-1654219 and OAC-1531814 and by the U.S. Department of Energy, Office of Science, Office of Nuclear Physics, within the framework of the Beam Energy Scan Theory (BEST) Topical Collaboration.
The authors acknowledges the use of the Maxwell Cluster and the advanced support from the Center of Advanced Computing and Data Systems at the University of Houston.

\appendix

\section{RHIC $v_n(p_T)$ and factorization breaking results} \label{sec:rhic}

\begin{figure}[ht]
\centering
\includegraphics[width=0.75\textwidth]{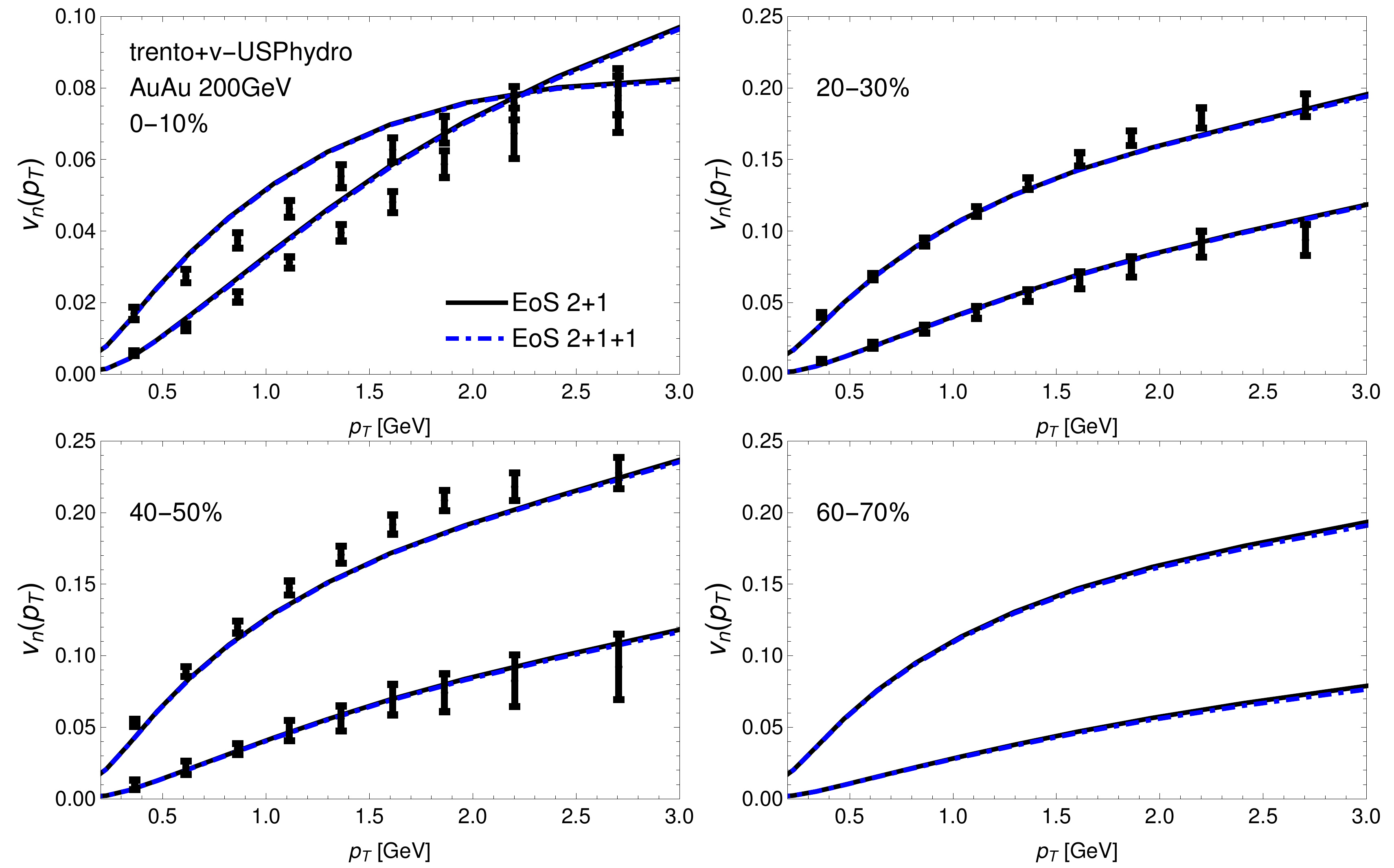}
\caption{(Color online) $v_n(p_T)$ calculations at AuAu $200$ GeV for charged particles for the 2+1-flavor EoS from \cite{Borsanyi:2013bia} and the 2+1+1-flavor EoS from \cite{Borsanyi:2016ksw}. }
\label{fig:vnptrhic}
\end{figure}

\begin{figure}[ht]
\centering
\includegraphics[width=0.75\textwidth]{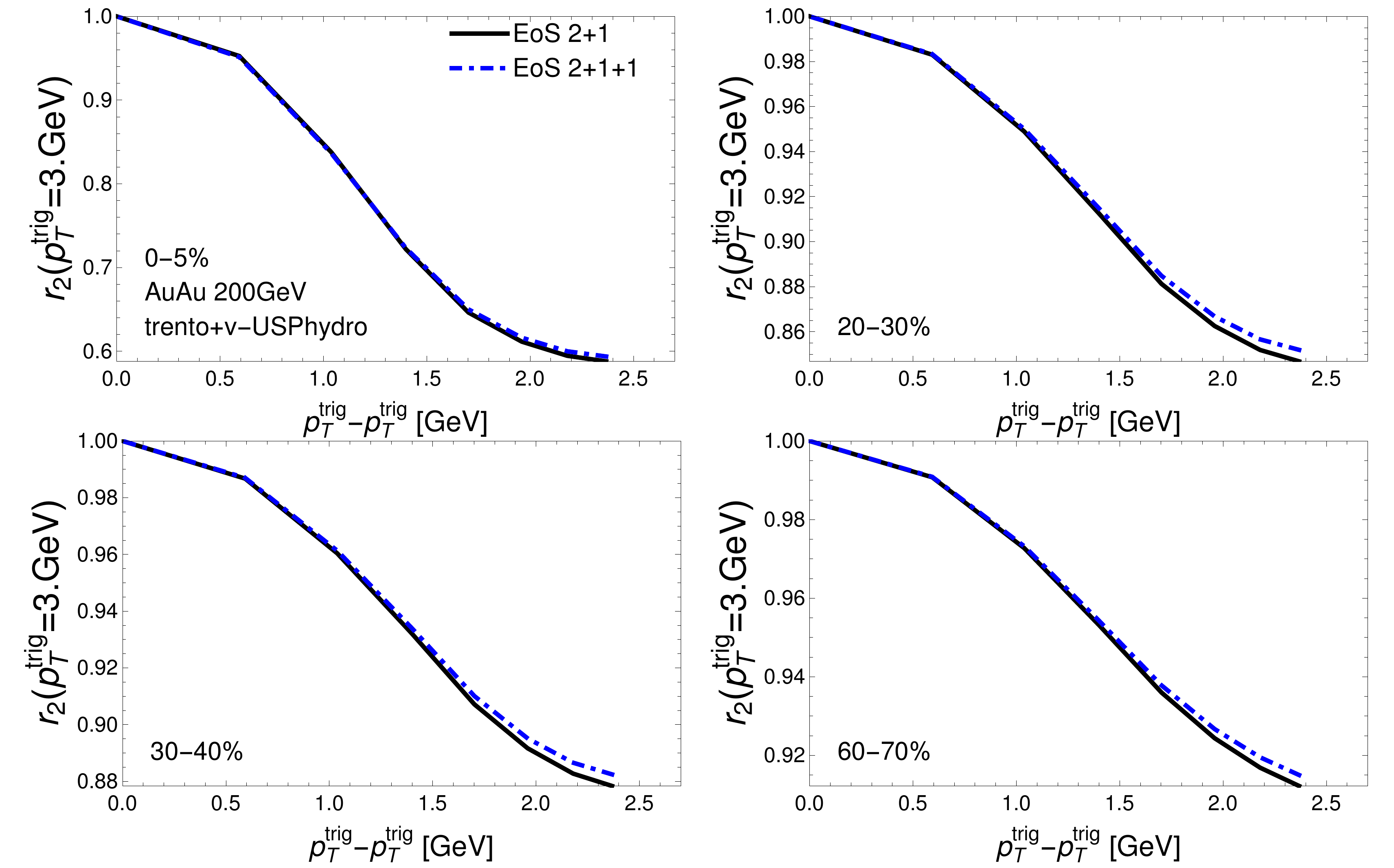}
\caption{(Color online) Factorization breaking at AuAu 200 GeV for charged particles for the 2+1-flavor EoS from \cite{Borsanyi:2013bia} and the 2+1+1-flavor EoS from \cite{Borsanyi:2016ksw}.}
\label{fig:rnrhic}
\end{figure}

The differences in the $v_n(p_T)$ and $r_2(p_T^a,p_T^b)$ at RHIC AuAu 200 GeV between the 2+1- and 2+1+1-flavor EoS are negligible, as shown in Fig. \ref{fig:vnptrhic} and Fig.\ \ref{fig:rnrhic}.  We note that both equations of state match experimental data points at STAR reasonably well in Fig.\ \ref{fig:vnptrhic} and we make predictions for the factorization breaking at RHIC in Fig.\ \ref{fig:rnrhic}.

\section{Note on event plane correlations}\label{sec:ep}

\begin{figure}[ht]
\centering
\includegraphics[width=0.75\textwidth]{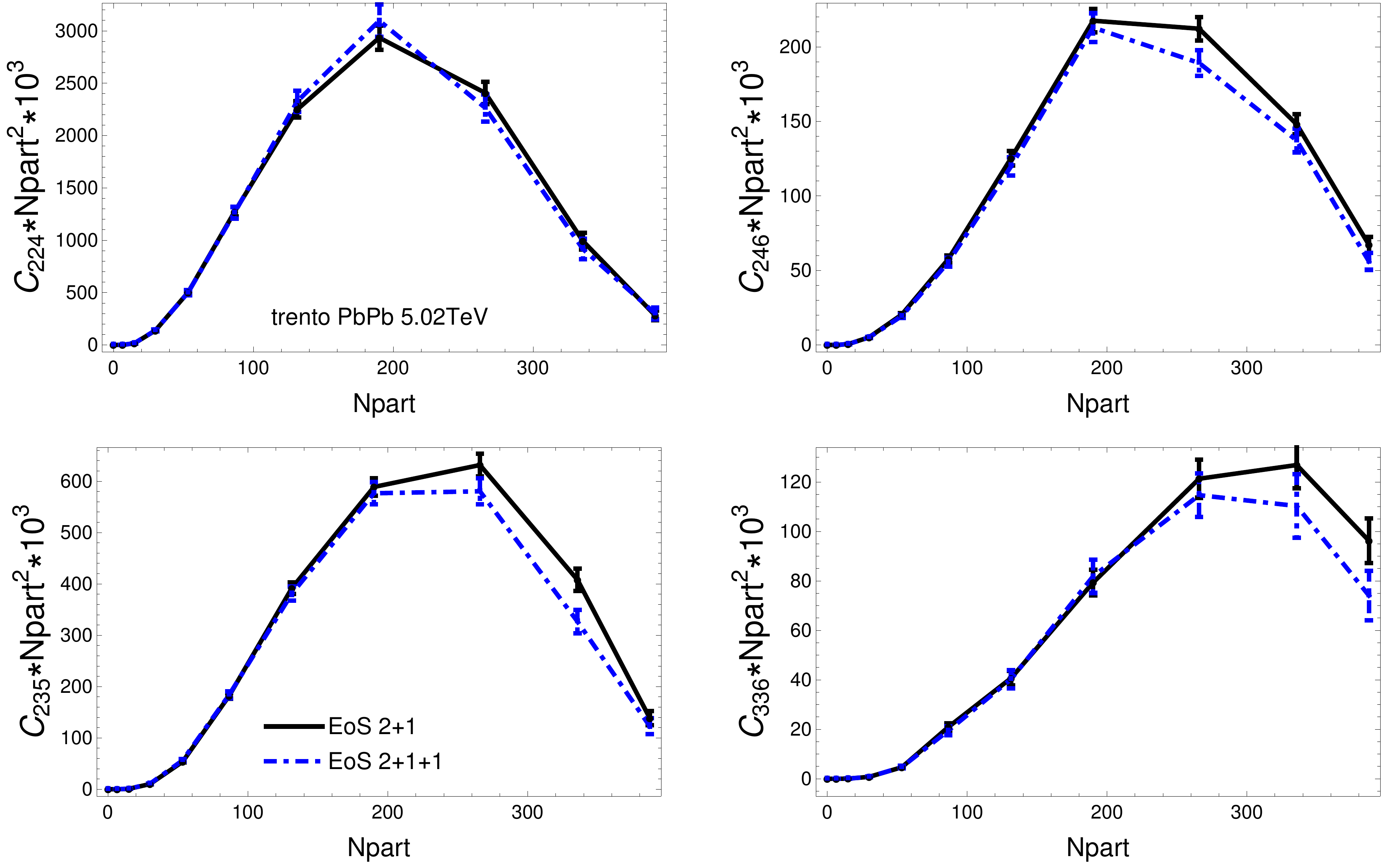}
\caption{(Color online) $Npart^2*C_{n,m,n+m}*10^3$ at PbPb $\sqrt{s_{NN}}=5.02$ TeV for charged particles for the 2+1-flavor EoS from \cite{Borsanyi:2013bia}, and the 2+1+1-flavor EoS from \cite{Borsanyi:2016ksw}. }
\label{fig:PbPb}
\end{figure}

\begin{figure}[ht]
\centering
\includegraphics[width=0.75\textwidth]{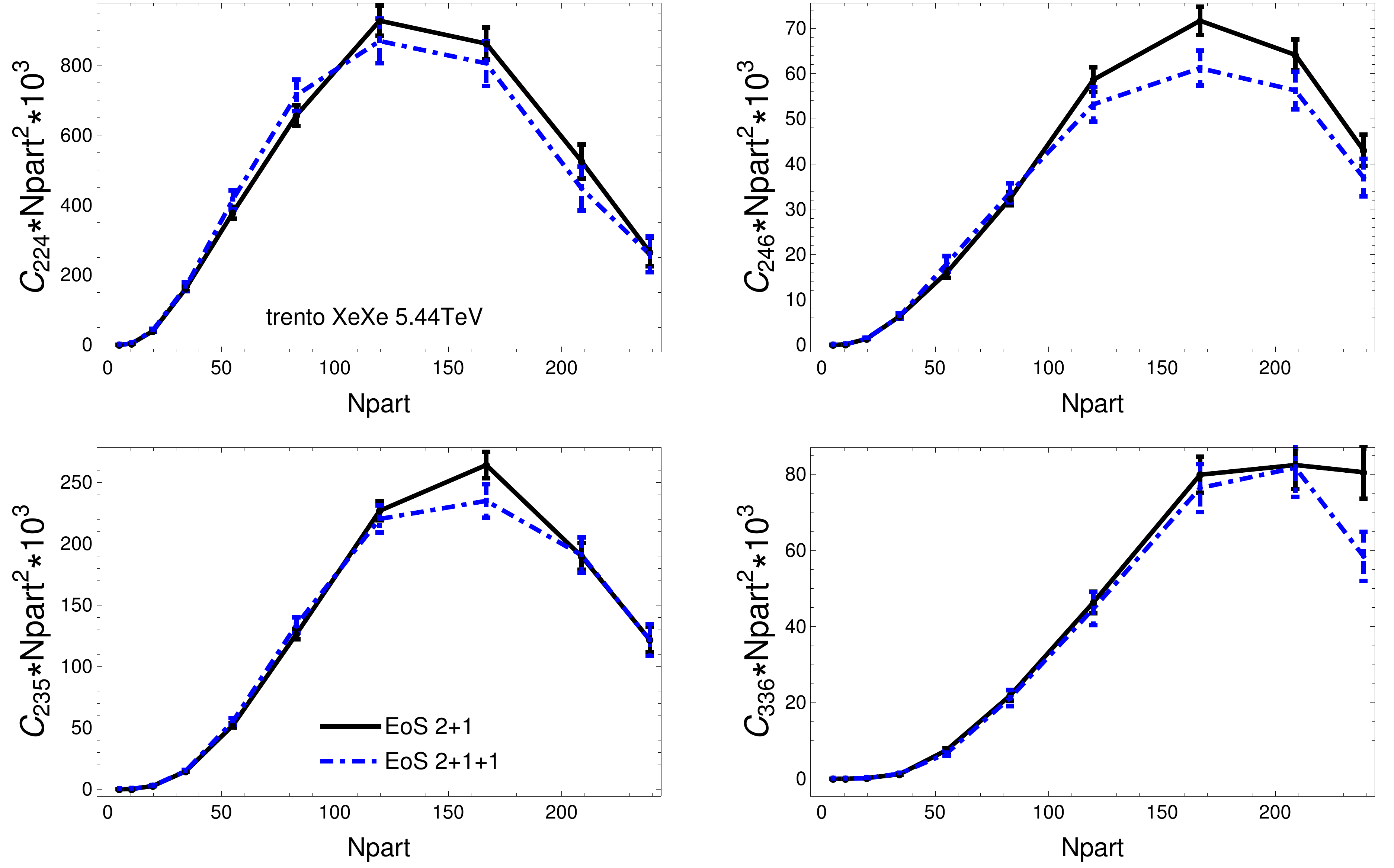}
\caption{(Color online) $Npart^2*C_{n,m,n+m}*10^3$ at XeXe $\sqrt{s_{NN}}=5.44$ TeV for charged particles for the 2+1-flavor EoS from \cite{Borsanyi:2013bia} and the 2+1+1-flavor EoS from \cite{Borsanyi:2016ksw}. }
\label{fig:XeXe}
\end{figure}

The event-plane correlations are calculated using the same method as from STAR \cite{Adamczyk:2017byf} at LHC PbPb 5.02 TeV in Fig.\ \ref{fig:PbPb} and at XeXe 5.44 TeV in Fig.\ \ref{fig:XeXe}. The differences between the two equations of state are small compared to those seen in the differential flow and factorization breaking. Only at RHIC energies are the differences in $Npart^2*C_{n,m,n+m}*10^3$ significant, since the EoS does not strongly affect the differential flow and factorization breaking.

\begin{figure}[ht]
\centering
\includegraphics[width=0.75\textwidth]{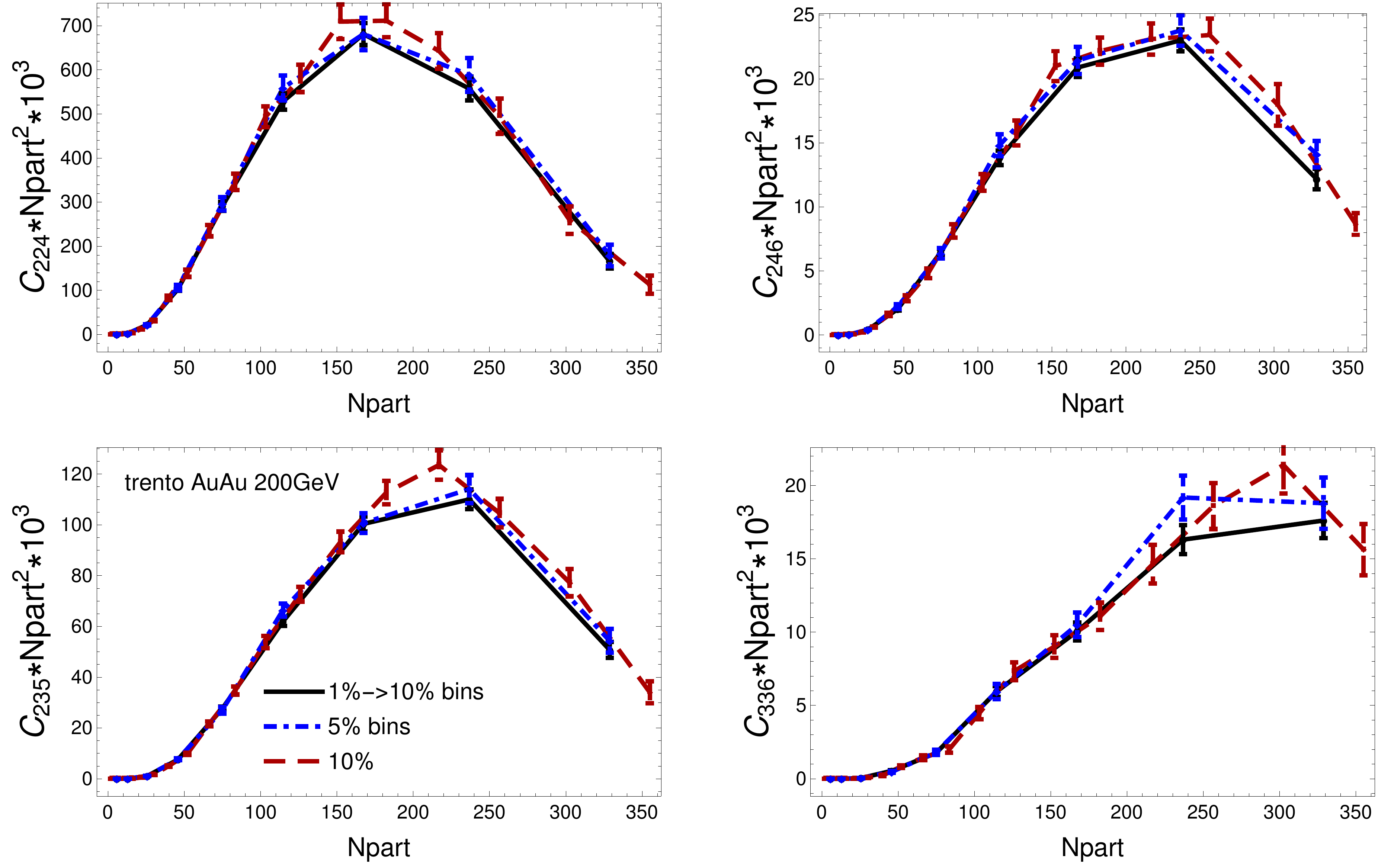}
\caption{(Color online) Effect of centrality binning widths on $Npart^2*C_{n,m,n+m}*10^3$ at RHIC for the 2+1 flavor EoS. }
\label{fig:binning}
\end{figure}

In a previous analysis at RHIC, it was found that the symmetric cumulants were affected by the multiplicity weighing and centrality rebinning \cite{Gardim:2016nrr}.  When one defines a centrality window of $10\%$, it may either be averaged over the entire centrality window or it could first be separated into $1\%$ centrality bins that are then later recombined into a $10\%$ .  The STAR measurements in \cite{Adamczyk:2017byf} did not implement centrality binning, so  in Fig.\ \ref{fig:binning} we explore how that would affect the final results. We find that $C_{336}$ is affected by the centrality binning in central collisions but generally the three different methods overlap in error bars. 

\section*{References}
\bibliography{all}{}

\end{document}